\documentclass[a4paper,12pt]{article}
\usepackage{amsmath,amssymb,amscd,cite}
\makeatletter
\renewcommand{\@seccntformat}[1]{\Roman{#1}.\hspace{.5em}}
\makeatother

\newfont{\zapfxii}{eusm10 scaled\magstep1}
\newcommand{\zap}[1]{\mbox{\zapfxii #1}}

\newcommand{\al}{\alpha}
\newcommand{\be}{\beta}
\newcommand{\ga}{\gamma}
\newcommand{\la}{\lambda}
\newcommand{\om}{\omega}

\newcommand{\zD}{{\zap D}}
\newcommand{\zT}{{\zap T}}

\def\bt{\bar{t}}
\def\bx{\bar{x}}
\def\bH{\overline{H}}
\def\bV{\overline{V}}
\def\bpsi{\bar\psi}

\newcommand{\NN}{{\mathbb N}}
\newcommand{\RR}{{\mathbb R}}

\def\pa#1{\partial_{#1}}

\def\abs#1{\left|#1\right|}

\newcommand{\im}{\operatorname{Im}}
\newcommand{\re}{\operatorname{Re}}

\newcommand{\dfo}{differential operator}
\newcommand{\sch}{Schr\"o\-ding\-er}
\newcommand{\QED}{\quad Q.E.D.}

\def\eg{e.g.}
\def\ie{i.e.}

\def\ni{\noindent}
\def\ns{\normalsize}
\def\ss{\small}

\def\sem{\ss\em}

\begin{document}
{\renewcommand{\thefootnote}{\fnsymbol{footnote}}
\begin{titlepage}
\begin{center}
\Large On form-preserving transformations\\
for the time-dependent \sch{} equation
\end{center}\vspace{.5cm}
\ns\textsc{Federico Finkel${}^{1}$, Artemio
Gonz\'alez-L\'opez${}^{1}$, Niky Kamran${}^{2}$,\\
Miguel A.~Rodr\'\i guez${}^{1}$}\\[.3cm]
{\sem ${}^1$ Departamento de F\'\i sica Te\'orica II, Universidad
Complutense de Madrid, \mbox{E-28040} Madrid, SPAIN\\
${}^2$ Department of Mathematics and Statistics, McGill University, Montr\'eal,
Qu\'ebec, CANADA H3A 2K6}\vspace{.2cm}

\center{\ns December 11, 1998}

\abstract{In this paper we point out a close connection between the
Darboux transformation and the group of point transformations
which preserve the form of the time-dependent \sch{} equation (TDSE).
In our main result, we prove that any pair of time-dependent real potentials
related by a Darboux transformation for the TDSE may be transformed
by a suitable point transformation into a pair
of time-independent potentials related by a usual Darboux
transformation for the stationary \sch{} equation.
Thus, any (real) potential solvable via a time-dependent
Darboux transformation can alternatively be solved by
applying an appropriate form-preserving point
transformation of the TDSE to a time-independent potential.
The preeminent role of the latter type of transformations
in the solution of the TDSE is illustrated with
a family of quasi-exactly solvable time-dependent anharmonic
potentials.}\vspace{.5cm}
\begin{flushleft}
\ni PACS numbers:\enspace 03.65.-w, 03.65.Ge

\ni {\em Running title:}\enspace
On form-preserving transformations for the TDSE
\end{flushleft}
\thispagestyle{plain}
\setcounter{footnote}{0}
\end{titlepage}}
\newpage
%
%
\setcounter{page}{2}
\section{Introduction}
A considerable amount of research has been devoted over the past few years
to the exact solution of the time-dependent \sch{} equation (TDSE) in $1+1$
dimensions. Several modifications of the celebrated Darboux transformation
for the stationary \sch{} equation, \cite{Da82, CKS95}, have been proposed
in this respect in the literature. Matveev and Salle showed that the usual
Darboux transformation for the stationary \sch{} equation could also be applied
to the TDSE with a time-dependent potential, \cite{MS91}. An equivalent
approach was followed by Bluman and Shtelen in~\cite{BS96}, who considered
a non-local transformation which is precisely the inverse map of the usual 
Darboux transformation. The Darboux transform of a time-dependent
potential is in general a complex-valued function. (The explicit conditions
for the resulting potential to be real-valued
appear in a recent paper by Bagrov and Samsonov,~\cite{BS96b}.)
For this reason, several generalizations of the Darboux transformation
mapping real potentials to real potentials have been proposed
in the literature. The best known of these generalizations is the
binary Darboux transformation described in \cite{MS91}, which is in fact
one of the main tools for finding exact solutions of integrable equations.

A seemingly unrelated method of constructing exact solutions of the TDSE
which has proved remarkably successful is based on the use of
point transformations which preserve the form of the TDSE. The idea goes
back to the work of Leach on the time-dependent harmonic oscillator, 
\cite{Le77}, arising, \eg, in the study of the motion
of charged particles in a Paul trap,~\cite{Br91}. The method
was subsequently extended by Bluman,~\cite{Bl80, Bl83},
and Ray,~\cite{Ra82}, to obtain exact solutions of the TDSE for a quadratic 
potential with arbitrary time-dependent coefficients. The technique
has also been applied to time-dependent
harmonic oscillators with a repulsive barrier, \cite{KP97},
and to anisotropic time-dependent harmonic potentials in
$2+1$ dimensions, \cite{BFFM79}.

The main purpose of this paper is to characterize the most general
time-dependent real potential whose Darboux transform is real. A partial result
along these lines for potentials rapidly decreasing at spatial infinity
was mentioned in \cite{BPPP91}. As noted in this reference, the
latter potentials are of limited interest regarded as solutions
of the KP equation.
However, potentials of this type (and, more generally, of the type considered
in this paper) are interesting from the point of view of exactly solving
the time-dependent Schr\"odinger equation, as underscored by the recent work of 
Bagrov and Samsonov, \cite{BS96b, BS97}.

In this paper we show that the Darboux transformation for the TDSE
is in fact closely related to the point transformation method. To this end,
in Section~\ref{sec.gen} we briefly review the Darboux transformation
and the point
transformations preserving the form of the TDSE, applying the latter
to construct a time-dependent anharmonic oscillator potential
admitting a certain number of algebraically computable wavefunctions. In
Section~\ref{sec.main} we derive the main results of our paper, proving that any
time-dependent real-valued potential for which the Darboux transformation yields
a real potential may always be mapped to a time-independent potential by a
form-preserving point transformation of the TDSE. Moreover, the Darboux
transformation for any such potential is equivalent to a Darboux transformation
for its associated time-independent potential, followed by the inverse of the
corresponding point transformation. Finally, Section~\ref{sec.con} is
devoted to our concluding remarks and related open questions.

\section{General background}\label{sec.gen}

In this section we summarize the basics of the Darboux and the form-preserving
point transformations for the TDSE. Following Bagrov and Samsonov, \cite{BS96b},
we take as the starting point of the Darboux transformation for the TDSE
the intertwining relation
\begin{equation}\label{intw}
L(i\pa t-H_0)=(i\pa t-H_1)L\,,
\end{equation}
where
\begin{equation}\label{Hsubi}
H_i=-\pa x^2+V_i(x,t)\,,\quad i=0,1\,,
\end{equation}
and $L$ is a first-order \dfo{} of the form
$$
L=L_1(x,t)\pa x+L_0(x,t)\,.
$$
It follows immediately from the intertwining
relation~\eqref{intw} that if $\psi_0$ verifies the TDSE with Hamiltonian
$H_0$, then $\psi_1=L\psi_0$ will solve the TDSE with Hamiltonian $H_1$.
It is also easily verified that the intertwining relation~\eqref{intw}
will be satisfied if and only if 
\begin{equation}\label{LandW}
L=L_{1}(\partial_{x}+\chi_{x})\,,\qquad
V_{1}=V_{0}+2\chi_{xx}+i(\log L_{1})_{t}\,,
\end{equation}
where $e^{-\chi}$ is a solution of the TDSE with potential $V_0$,
and $L_1=L_1(t)$ is an arbitrary function. The transformed potential
$V_1(x,t)$ is a real-valued function if and only if
\begin{align}
& \im\chi_{xxx}=0\label{thecon}\\
\intertext{and}
& |L_1|=\exp\left[-2\int^t_{t_0} \im\chi_{xx}(x,s)\,ds\right]\,.
\label{modL1}
\end{align}
Without loss of generality, we shall assume from now on that $L_1$
is real and positive, and is therefore given by the right-hand side
of~\eqref{modL1}.

Just as in the time-independent case, the Darboux transformation
for the TDSE can be inverted. Indeed, if $\psi_1$ is a solution of the TDSE with
potential $V_{1}$ given by~\eqref{LandW}, the function
\begin{equation}\label{invmap}
\psi_0(x,t)=\frac{e^{-\chi(x,t)}}{L_1(t)}\left[\int_{x_0}^x
e^{\chi(y,t)}\psi_1(y,t)dy+ c_0(t)\right]
\end{equation}
with $c_0(t)$ given by\\
$$
c_0(t)=iL_1(t)\int_{t_0}^t \frac{e^{\chi(x_0,s)}}{L_1(s)}
\big(\psi_{1,x}(x_0,s)-\chi_x(x_0,s)\psi_1(x_0,s)
\big)\,ds
$$
verifies the TDSE with potential $V_0$. If the factor $L_1$ is taken
as unity, the mapping $\psi_1\mapsto\psi_0$ given by~\eqref{invmap}
reduces to the non-local transformation
considered by Bluman and Shtelen in \cite{BS96}.\\

The most general point transformation mapping the TDSE
\begin{equation}\label{TDSE}
\big(i\partial_{t}+{\partial_{x}}^2-V_{0}(x,t)\big)\psi_{0}(x,t)=0
\end{equation}
for any given potential $V_0$ into another TDSE with potential
$\bV_0$ for the transformed wavefunction $\bpsi_0$ is defined by
\begin{gather}
\bx=\frac x{C(t)}+B(t)\,,\qquad\qquad
\bt=\int^t_{t_0}\frac{ds}{C^2(s)}\,,\notag\\
\psi_0(x,t)=|C|^{-1/2}
\exp\bigg[\frac i4\bigg(\frac{\dot C}C x^2-2\dot BC x+A(t)\bigg)\bigg]
\bpsi_0(\bx,\bt)\,,\label{br}\\
V_0(x,t)=\frac 1{C^2}\,\bV_0(\bx,\bt)-\frac{\ddot C}{4C}\, x^2
+\Big(\frac{C\ddot B}2+\dot B\dot C\Big)x
-\frac 14\big(C^2\dot B^2+\dot A\big)\,,\notag
\end{gather}
where $A$, $B$ and $C\neq 0$ are real-valued functions of $t$.
Note that square-integrability is preserved under the
transformation~\eqref{br}. As remarked before, the
point transformation~\eqref{br}
has been employed to construct exact solutions
of the TDSE for quadratic potentials with time-de\-pend\-ent coefficients.
The interest of the transformation~\eqref{br} is
not limited, however, to quadratic (or exactly solvable)
time-dependent potentials, as evidenced by the following\\

{\ni\bf Example.} Consider the two-parameter family of anharmonic oscillator
potentials given by, \cite{SBD78, TU87, GKO93},
\begin{equation}\label{sextic}
\bV(\bx)=\bx^6+2\al\bx^4+(\al^2-4n-3)\bx^2\,,
\end{equation}
where $\al\in\RR$ and $n\in\NN$. The sextic potential~\eqref{sextic} is a
well-known example of the class of {\em quasi-exactly solvable} potentials,
for which a certain subset of the spectrum can be computed by purely
algebraic means; see~\cite{Us94} for an extensive review of the field.
The first $n+1$ even bound states of the potential~\eqref{sextic}
are of the form
\begin{equation}\label{states}
\phi(\bx)=\exp\left[-\frac 14\bx^4-\frac\al 2 \bx^2\right]p(\bx^2)\,,
\end{equation}
where $p(s)$ is a polynomial in $s$ of degree less than or equal to $n$ which
can be computed algebraically. The point transformation~\eqref{br} with
$A=B=0$ and
$C=\om^{-1/2}$, $\om=\om(t)$ being a positive function,
leads directly to the potential $V(x,t)$ given by
\begin{equation}\label{tsextic}
V(x,t)=\om^4 x^6+2\al\,\om^3 x^4
+\Big(\al^2-4n-3-{3\dot \om^2-2\om\ddot \om\over{16 \om^4}}\Big)\om^2 x^2\,.
\end{equation}
The TDSE with potential~\eqref{tsextic} possesses $n+1$ square-integrable
solutions of the form
\begin{equation}\label{sols}
\psi(x,t)=\om^{1/4}\exp\Big[-i\Big(\frac{\dot \om}{8\om}\, x^2
+E\int_{t_0}^t \om(s)ds\Big)\Big]\phi\big(\sqrt{\om}\, x\big)\,,
\end{equation}
where $\phi(\bx)$ is an algebraic eigenfunction of the form~\eqref{states}
with eigenvalue $E$ of the Hamiltonian
$$
\bH=-\pa\bx^2+\bV(\bx)\,.
$$
The potential~\eqref{tsextic} thus provides a natural extension
of the notion of quasi-exact solvability to the time-dependent case,
in the sense that the associated TDSE admits a certain number of
solutions which can be determined algebraically. In particular, note
that if $\om(t)$ is of the form
$$
\om_0(t)=\be\left[\ga+\big(\ga^2-(\al^2-4n-3)\be\big)^{1/2}
\sin\big(4\sqrt\be t+\delta\big)\right]^{-1}\,,
$$
with $\delta\in\RR$, $\be>0$ and $\ga^2>(\al^2-4n-3)\be>0$, the
potential~\eqref{tsextic} reduces to a harmonic oscillator with a
periodic-in-time anharmonic perturbation, namely
$$
V_0(x,t)=\om_0^4(t) x^6+2\al\,\om_0^3(t) x^4+\be x^2\,.
$$

\section{The reality condition and the form-preserving point
transformations}
\label{sec.main}

In this section we prove the main results of our paper, starting with the
following\\

{\ni\bf Theorem.} Let $e^{-\chi}$ be a solution of the TDSE with
potential $V_0(x,t)$. If $\chi$ satisfies the reality
condition~\eqref{thecon}, then $V_0(x,t)$ may be mapped
to a time-independent potential $\bV_0(\bx)$ by a point
transformation~\eqref{br}.\\
{\ni\em Proof.} Let $\chi_0=\re\chi$, $\chi_1=\im\chi$. The
TDSE for $e^{-\chi}$ is then equivalent to the pair of real equations
given by
\begin{align}
& \chi_{0,t}+\chi_{1,xx}-2\chi_{0,x}\chi_{1,x}=0\,,\label{im}\\
& V_0(x,t)=\chi_{1,t}-\chi_{0,xx}-\chi_{1,x}^2+\chi_{0,x}^2\,.\label{re}
\end{align}
If the reality condition~\eqref{thecon} holds, \ie, if
$\chi_1$ is of the form
\begin{equation}\label{chi1}
\chi_1=a(t)x^2+b(t)x+c(t)\,,
\end{equation}
eqns.~\eqref{im} and~\eqref{re} reduce to
\begin{align}
& \chi_{0,t}-2(2 a\,x+b)\chi_{0,x}+2a=0\,.\label{im2}\\
& V_0(x,t)=(\dot a-4 a^2)x^2+(\dot b-4ab)x+\dot c-b^2+\chi_{0,x}^2-\chi_{0,xx}
 \label{re2}\,.
\end{align}
The general solution of eq.~\eqref{im2} is of the form \cite{fn1}
\begin{equation}\label{chi0}
\chi_0=-2\int_{t_0}^t a(s)ds+F\left(
e^{4\int_{t_0}^t a(s)ds}\,x+2\int_{t_0}^t b(s)e^{4\int_{t_0}^s
a(r)dr}ds\right)\,,
\end{equation}
where $F$ is an arbitrary real-valued function.
Substituting this expression into~\eqref{re2}, we immediately conclude
that the transformation~\eqref{br} determined by
\begin{equation}\label{CBA}
C(t)=e^{-4\int_{t_0}^t a(s)ds}\,,\qquad
B(t)=2\int_{t_0}^t\frac{b(s)}{C(s)}\,ds\,,
\qquad A(t)=-4c(t)\,,
\end{equation}
maps the potential $V_0(x,t)$ into the time-independent potential
$$
\qquad\qquad\ \bV_0(\bx)={F'}^2(\bx)-F''(\bx)\,.\qquad\QED
$$

Thus, any potential for which the
Darboux transformation yields another real-valued potential may be mapped
into a time-independent potential by a form-preserving point transformation.
It is easy to check that the transform of $e^{-\chi}$
under the point transformation~(\ref{br},\ref{chi1},\ref{chi0},\ref{CBA})
is $e^{-F}$, which is therefore an eigenfunction of the time-independent
potential $\bV_0(\bx)$. The next Corollary shows that
the usual Darboux transform of the associated time-independent potential
generated by $F$ is related to the Darboux transform of the original potential
by the same point transformation:\\

{\ni\bf Corollary.} Let $e^{-\chi}=e^{-\chi_0-i\chi_1}$ be a solution
of the TDSE with potential $V_0(x,t)$, with $\chi$ satisfying the reality
condition~\eqref{thecon}. Let $\zD$ and $\zT$ denote, respectively,
the Darboux transformation~\eqref{LandW} and the point
transformation~\eqref{br} defined by $\chi$ via
eqns.~(\ref{chi1},\ref{chi0},\ref{CBA}).
Let $\overline\zD$ denote the Darboux transformation
generated by $F$. Then
\begin{equation}
\label{darbouxray}
\zT\circ\zD=\overline\zD\circ\zT\,.
\end{equation}
{\ni\em Remark.} This result may be easily visualized with the help of the
following commutative diagram:
$$
\begin{CD}
V_0(x,t)@>\zT>>\bV_0(\bx)={F'}^2-F''\\
@V\zD VV @VV\overline\zD V\\
V_1(x,t)@>>\zT>\bV_1(\bx)={F'}^2+F''
\end{CD}
$$
{\ni\em Proof.} The proof follows from a straightforward application
of the appropriate formulae for the transformed potentials and
wavefunctions.\QED\\

The above Corollary shows, in particular, that the potential $V_1(x,t)$ is
the image under the inverse of the form-preserving point transformation $\zT$
determined by (\ref{br},\ref{chi1},\ref{chi0},\ref{CBA}) of a time-independent
potential $\bV_1(\bx)$. Exact solutions of the TDSE with potential $V_1$ can
therefore be obtained simply by applying the point transformation
$\zT\kern1pt^{-1}$ to solutions of the TDSE for the {\em time-independent}
potential $\bV_1$.

Another important consequence of the above Corollary is that, if the
potential $\bV_0$ satisfies (for instance) the condition
\begin{equation}
\label{l11}
\int_{-\infty}^\infty\abs{\bV_0(\bx)}(1+\abs\bx)\,d\bx<\infty\,,
\end{equation}
then the time-de\-pendent Darboux transformation~\eqref{LandW} preserves the
square-integrability of eigenfunctions. Indeed, if $\bV_0$ verifies \eqref{l11}
then the time-independent Darboux transformation $\overline\zD$ determined by a
non-vanishing eigenfunction of $\bV_0$ preserves square integrability,
\cite{DT79}, \cite{SB95}. The result stated above then follows easily from
\eqref{darbouxray}, the invertibility of $\zT$, and the fact that the
form-preserving point transformation $\zT$ always preserves square
integrability.\\

{\ni\bf Example.} 
It is straightforward to verify that all the examples of time-dependent
potentials appearing in Refs.~\cite{BS96, BS96b, BS97} which are solvable by
means of a Darboux transformation are indeed the images of certain exactly
solvable time-independent potentials under suitable form-preserving point
transformations. 

For instance, the free-particle potential $V_0(x,t)=0$ admits
a one-parameter family of solutions
$$
\psi_\la(x,t)=(1+t^2)^{-1/4}\exp\left[\frac i4\left(
\frac{t\,x^2}{1+t^2}+4\,\la\,\arctan
t\right)\right]Q_\la(x/\sqrt{1+t^2})
$$
satisfying the reality condition \eqref{thecon}, where $Q_\la$ is a
(real-valued) solution of Weber's equation
\begin{equation}
\label{weber}
Q_\la''(y)-\left(\frac{y^2}4+\la\right)\,Q_\la(y)=0
\end{equation}
(see \cite{BS97}, eq.~(18)).
By the Theorem at the beginning of this section, it follows that $V_0$
is related to a certain time-independent potential $\bV_0$ by a point
transformation~\eqref{br}. Indeed, in this case we have
$$
\chi_1=-\im\log\psi_\la=-\frac{t\,x^2}{4(1+t^2)}-\la\arctan
t\,,
$$
so from \eqref{CBA} it follows that
\begin{equation}
\label{pointt}
C=\sqrt{1+t^2},\qquad B=0,\qquad A=4\,\la\,\arctan t.
\end{equation}
Substituting these formulae into \eqref{br} we find that
\begin{equation}
\label{bxex}
\bx = \frac x{\sqrt{1+t^2}}
\end{equation}
and
\begin{equation}
\bV_0(\bx)=\frac{\bx^2}4+\la
\end{equation}
is a harmonic oscillator potential.

For all $n\in\NN$, let $H_n$ denote the $n$-th Hermite polynomial. The
functions \cite{fn2}
$$
Q_{n+\frac12}(y)=i^n\,e^{y^2/4}H_n(i\,y/\sqrt2)
$$
are real-valued solutions of Weber's equation \eqref{weber} with $\la=n+\frac12$
without zeros on the positive real semiaxis \cite{fn3}. Hence, for
all $n\in\NN$ the Darboux
transformation determined by the eigenfunction
$\psi_{n+\frac12}$ of $V_0$ is well-defined on the positive real semiaxis.
From eqns.~(\ref{LandW},\ref{modL1}) and the definition of
$\psi_\la$, it follows that the transformed potential
$V_1$ is given by
$$
V_1(x,t)=2\chi_{0,xx}=-2\left[\log Q_{n+\frac12}(\bx)\right]_{xx}
=\frac2{1+t^2}\left(
\frac{Q'{}^2_{n+\frac12}(\bx)}{Q_{n+\frac12}^2(\bx)}
- \frac{\bx^2}4-n-\frac12\right),
$$
where $\bx$ is given by \eqref{bxex}. Using standard identities for the
derivatives of the Hermite polynomials, the reader can easily verify that this
formula for
$V_1(x,t)$ agrees with the corresponding expression given in \cite{BS97}.
As stated in the Corollary, the potential $V_1(x,t)$ is related by the point
transformation \eqref{br} defined by \eqref{pointt} to a time-independent
potential $\bV_1(\bx)$ obtained from $\bV_0(\bx)$ by applying a
time-independent Darboux transformation
$\overline\zD$. {}From \eqref{chi0} and the definition of $\psi_\la$, it
easily follows that the function $F(\bx)$ generating the Darboux
transformation $\overline\zD$ is given by
$$
F(\bx)=-\log Q_{n+\frac12}(\bx)\,,
$$
and therefore
$$
\bV_1(\bx)=F'{}^2(\bx)+F''(\bx)=
2\frac{Q'{}^2_{n+\frac12}(\bx)}{Q_{n+\frac12}^2(\bx)}
- \frac{\bx^2}4-n-\frac12\,.
$$
It is straightforward to check that the potentials $V_1(x,t)$ and $\bV_1(\bx)$
are indeed related by the point transformation \eqref{br} determined by
\eqref{pointt}.

\section{Conclusions}\label{sec.con}

In this paper, we have shown that the Darboux transformation for
the time-dependent \sch{} equation is essentially equivalent
to the usual Darboux transformation for the stationary \sch{} equation.
Any (real) potential $V_1(x,t)$ solvable via a time-dependent
Darboux transformation starting from a real potential $V_0(x,t)$
can alternatively be solved by applying a form-preserving point
transformation to a time-independent potential $\bV_1(\bx)$.
As a matter of fact, although a large number of methods
and papers have been devoted in recent times to the exact solution of the
TDSE \cite{fn4}, most of the associated
time-dependent Hamiltonians either are not of the standard form~\eqref{Hsubi},
or are also obtainable from a time-independent Hamiltonian by a form-preserving
point transformation.

The interest of the Darboux transformation for the TDSE
as a useful method to obtain new (quasi-)exactly solvable time-dependent
potentials is therefore very limited. It should be noted, however, that
the Darboux transformation for the TDSE could still render helpful results if
the starting potential is not a real-valued function but only the
transformed potential is real \cite{fn5}. As a final remark, we would like to stress that
the Darboux transformation may still be useful to construct exact solutions to
real-valued diffusion equations of the Fokker--Planck type,
for which no reality condition as eq.~\eqref{thecon} must be considered.

\section*{Acknowdgements}
FF, AG-L, and MAR would like to acknowledge the partial financial support of the
DGICYT under grant no.~PB95--0401. NK was supported in part by NSERC
grant \#0GP0105490.


\begin{thebibliography}{99}
\bibitem{fn1} Note that
eq.~\eqref{im} cannot in general be solved in closed form
if $\chi_1$ is a polynomial in $x$ of degree $k>2$.
\bibitem{fn2} Our choice of $Q_{n+\frac12}$ differs from the one in
\cite{BS97} by the irrelevant constant factor $i^n$.
\bibitem{fn3} If $n$ is even
$Q_{n+\frac12}$ is non-zero on the whole real axis.
\bibitem{fn4} A comprehensive review of these methods
is beyond the scope of this paper; see~\cite{Ma54, WN63, Tr81, CF88, ZL96}
and references therein for a detailed treatment.
\end{thebibliography}
\end{document}